\documentstyle[12pt,a4wide,pstcol,epsf]{article}

\newcommand{\calp}{{\cal P}}

\begin{document}
\begin{center}
{\LARGE Shaft Inflation}

\medskip

{\bf Konstantinos Dimopoulos}

{\em Physics Department, Lancaster University, Lancaster LA1 4YB, UK}%
\footnote{e-mail: {\tt k.dimopoulos1@lancaster.ac.uk}}
\end{center}


\begin{abstract}
A new family of inflation models is introduced and studied. The models are
characterised by a scalar potential which, far from the origin, approximates an
inflationary plateau, while near the origin becomes monomial, as in chaotic 
inflation. The models are obtained in the context of global supersymmetry 
starting with a superpotential, which interpolates from a generalised monomial 
to an O'Raifearteagh form for small to large values of the inflaton field 
respectively. It is demonstrated that the observables obtained, such as the 
scalar spectral index and the tensor to scalar ratio, are in excellent 
agreement with the latest observations. Some discussion of 
initial conditions and eternal inflation is included.
\end{abstract}

\bigskip


The latest CMB observations from the Planck satellite have confirmed the
broad predictions of the inflationary paradigm, in that the Universe is
found to be spatially flat with a predominantly Gaussian curvature perturbation
that is almost (but not quite) scale invariant \cite{planckinf}. However, the 
precision of these observations is so high that they put tension to (or even 
exclude) entire classes of inflationary models, e.g. chaotic inflation.%
\footnote{unless excited states are assumed instead of the usual Bunch-Davis 
vacuum \cite{ash}.} 

The Planck observations seem to support an inflationary scalar potential which 
asymptotes to a constant, i.e. an inflationary plateau is favoured 
\cite{bestinf}. In view of this fact, in this letter we present a new class of 
inflationary potentials, which we call shaft inflation. The idea is that the 
inflationary plateau is pierced by shafts such that, when the inflaton field 
finds itself close to one of them it slow-rolls inside the shaft, until 
inflation ends and gives away to the hot big bang cosmology. Assuming a shaft 
at the origin, the scalar potential approximates a constant at large values of 
the inflaton field, 
but at small values the 
potential becomes similar to monomial chaotic inflation. In that respect, shaft 
inflation is similar to the so-called T-model inflation \cite{Tmodel} but the 
scalar potential in our case features a power-law (in contrast to exponential)
dependence on the inflaton field.
Although we attempt to design the model in the context of global supersymmetry,
this is by no means restrictive since the phenomenology really stems out from 
the form of the scalar potential, which can be obtained via a different, 
possibly more realistic (and complicated) setup. Indeed, as we discuss, one of 
the realisations of shaft inflation can be identified with S-dual superstring 
inflation \cite{Sdualinf} or with radion assisted gauge inflation~\cite{RAGI}. 

After the first draft of this letter was produced, the first data of the BICEP2
experiment were released, which show that inflation may produce substantial 
gravitational waves. According to the findings of BICEP2, the tensor to scalar 
ratio is \mbox{$r\simeq 0.20\pm 0.07$} \cite{bicep2}. We show that shaft 
inflation can accommodate such a large value of $r$.

We use natural units, where \mbox{$c=\hbar=1$} and Newton's gravitational 
constant is \mbox{$8\pi G=m_P^{-2}$}, with \mbox{$m_P=2.43\times 10^{18}\,$GeV} 
being the reduced Planck mass.

Let us begin with the following superpotential:
\begin{equation}
W=M^2\frac{|\phi|^{nq+1}}{(|\phi|^n+m^n)^q}
\label{W0}
\end{equation}
where $M,m$ are mass-scales, $n,q$ are real parameters and $\phi$ is a real 
scalar field (corresponding to a superfield made real by suitable field 
redefinitions). Without loss of generality, we assume that \mbox{$\phi>0$}
so we can write \mbox{$|\phi|=\phi$} and assume that there is a $Z_2$ symmetry
\mbox{$\phi\rightarrow-\phi$}. In the limit \mbox{$\phi\gg m$} the above 
superpotential reduces to an O'Raifearteagh form \mbox{$W\simeq M^2\phi$},
which leads to de-Sitter expansion. However, in the limit \mbox{$\phi\ll m$}
the superpotential becomes \mbox{$W\simeq M^2\phi(\phi/m)^{nq}$}, which leads to
monomial chaotic inflation. To simplify the potential we may assume 
\mbox{$q=-1/n$}, in which case the superpotential becomes
\begin{equation}
W=M^2\left(\phi^n+m^n\right)^{1/n},
\label{W}
\end{equation}
Thus, in the limit \mbox{$\phi\ll m$} the above becomes
\mbox{$W\simeq M^2m+\frac{1}{n}M^2m(\phi/m)^n$}, which leads to a monomial
F-term potential. 
\footnote{The form of the superpotential in Eq.~(\ref{W0}) is dictated by the 
requirement that it gives rise to the envisaged scenario. The simplifying 
relation \mbox{$q=-1/n$} reduces the parameters and can have physical 
interpretation for specific values of \mbox{$n$}, e.g. \mbox{$n=2$} (see below).
However, the physical meaning of Eq.~(\ref{W}) falls beyond the scope of this 
letter; while in a sense, can be thought as the definition of shaft inflation.}

\begin{center}
\begin{figure}

\vspace{20cm}

\begin{picture}(10,10)
\put(-20,0){
\leavevmode
\hbox{\epsfxsize=7in
\epsffile{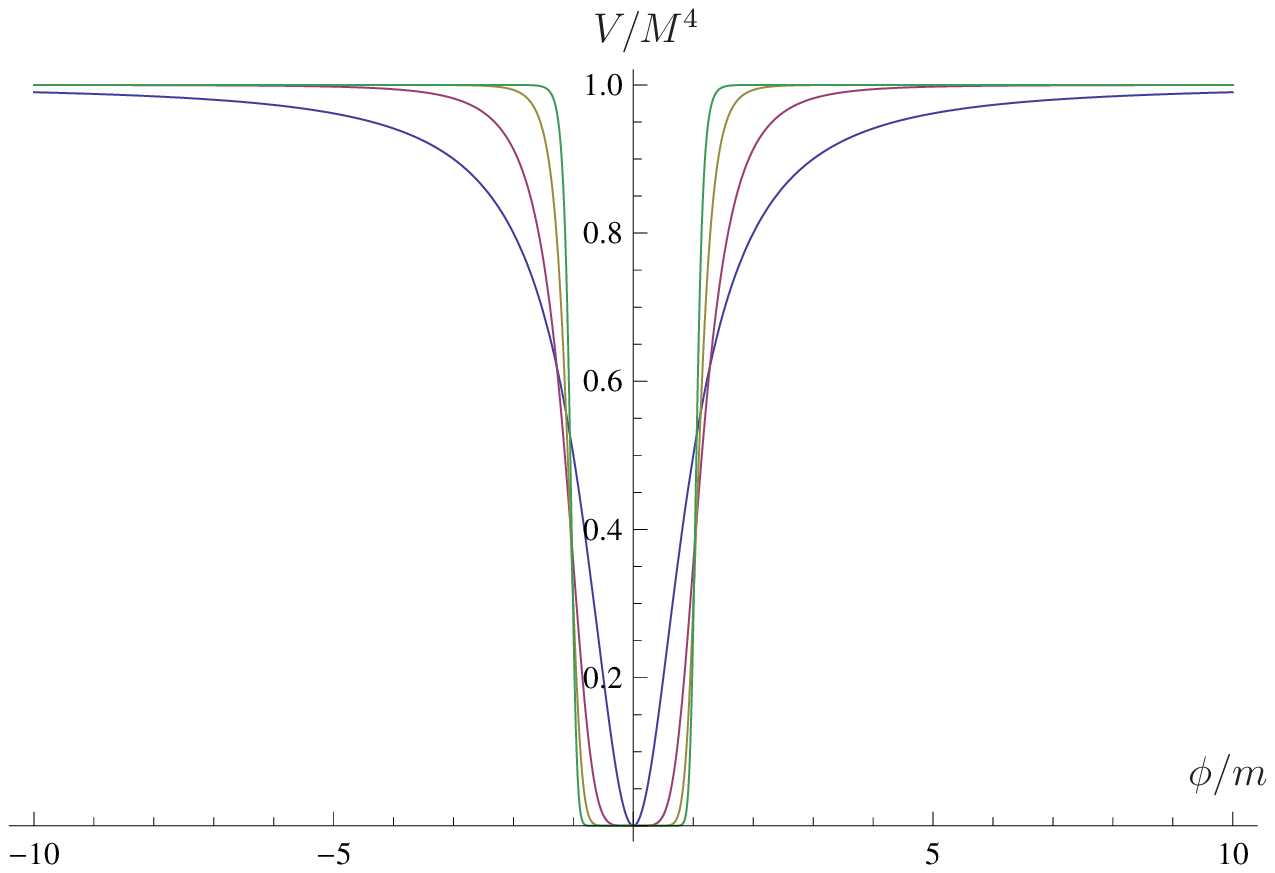}}
}
\end{picture}
\vspace{-13cm}
\caption{\footnotesize
The scalar potential in shaft inflation for \mbox{$n=2,4,8$} and 16. The shaft 
becomes sharper as $n$ grows. Far from the origin the potential approximates 
the inflationary plateau with \mbox{$V\approx M^4$}. Near the origin 
the potential becomes monomial, as in chaotic inflation. 
}
\label{fig1}


\end{figure}
\end{center}

For the superpotential in Eq.~(\ref{W}), the corresponding F-term scalar 
potential is:
\begin{equation}
V(\phi)=
M^4\phi^{2n-2}(\phi^n+m^n)^{\frac{2}{n}-2}.
\label{V}
\end{equation}
From the above we see that the scalar potential has the desired behaviour
for \mbox{$n>1$}, i.e. it approaches a constant for \mbox{$\phi\gg m$},
while for \mbox{$\phi\ll m$} the potential becomes monomial, with
\mbox{$V\propto\phi^{2(n-1)}$}, see Fig.~\ref{fig1}. When \mbox{$n=1$} the 
potential is exactly flat and the shaft disappears. 


\vspace{-1cm}

For the slow-roll parameters we find
\begin{eqnarray}
& & \epsilon\equiv\frac12 m_P^2\left(\frac{V'}{V}\right)^2=
2(n-1)^2\left(\frac{m_P}{\phi}\right)^2\left(\frac{m^n}{\phi^2+m^n}\right)^2
\label{eps}\\
& & \eta\equiv m_P^2\frac{V''}{V}= 
2(n-1)\left(\frac{m_P}{\phi}\right)^2\left(\frac{m^n}{\phi^2+m^n}\right)
\frac{(2n-3)m^n-(n+1)\phi^n}{\phi^n+m^m},
\label{eta}
\end{eqnarray}
where the prime denotes derivative with respect to the inflaton field. Hence, 
the spectral index of the curvature perturbation is
\begin{equation}
n_s=1+2\eta-6\epsilon=1-4(n-1)\left(\frac{m_P}{\phi}\right)^2
\frac{m^n[(n+1)\phi^n+nm^n]}{(\phi^n+m^n)^2}.
\label{ns}
\end{equation}

It is straightforward to see that inflation is terminated when 
\mbox{$|\eta|\simeq 1$} so that, for the end of inflation, we find
\begin{equation}
\phi_{\rm end}\simeq m_P\left[2(n^2-1)\alpha^n\right]^{1/(n+2)},
\label{fend}
\end{equation}
where we assumed that \mbox{$\phi>m$} (so that the potential deviates
from a chaotic monomial) and we defined
\begin{equation}
\alpha\equiv\frac{m}{m_P}\,.
\label{alpha}
\end{equation}
Using this, we obtain $\phi(N)$ 
\begin{eqnarray}
N=\frac{1}{m_P^2}\int_{\phi_{\rm end}}^\phi\frac{V}{V'}{\rm d}\phi & \simeq &
\frac{1}{2(n-1)(n+2)\alpha^n}\left[\left(\frac{\phi}{m_P}\right)^{n+2}-
\left(\frac{\phi_{\rm end}}{m_P}\right)^{n+2}\right]
\label{N}\\
& \Rightarrow & \phi(N)\simeq m_P\left[2(n-1)(n+2)\alpha^n
\left(N+\frac{n+1}{n+2}\right)\right]^{1/(n+2)},
\label{fN}
\end{eqnarray}
where $N$ is the remaining e-folds of inflation and we considered 
\mbox{$\phi>m$} again. Inserting the above into Eqs.~(\ref{eps}) and (\ref{ns})
respectively we obtain the tensor to scalar ratio $r$ and the spectral index 
$n_s$ as functions of $N$:
\begin{eqnarray}
& & r=16\epsilon=32(n-1)^2\alpha^{\frac{2n}{n+2}}\left[2(n-1)(n+2)
\left(N+\frac{n+1}{n+2}\right)\right]^{-2(\frac{n+1}{n+2})}
\label{r}\\
& & n_s=1-2\frac{n+1}{n+2}\left(N+\frac{n+1}{n+2}\right)^{-1}
\label{nsN}
\end{eqnarray}

An interesting choice is \mbox{$n=2$}, 
in which case the scalar potential becomes
\begin{equation}
V(\phi)=M^4\frac{\phi^2}{\phi^2+m^2}.
\label{Vquad}
\end{equation}
We see that the above can be thought of as a modification of quadratic 
chaotic inflation, because after the end of inflation, the inflaton field 
oscillates in a quadratic potential. However, for large values of the inflaton 
the potential approaches a constant. 
This potential has been obtained also in S-dual superstring inflation 
\cite{Sdualinf} with \mbox{$\alpha=1/4$} and also in radion assisted gauge 
inflation \cite{RAGI} with \mbox{$\alpha\sim 10^{-3/2}$}. In this case,
Eqs.~(\ref{r}) and (\ref{nsN}) become
\begin{equation}
r=\frac{32\alpha}{\left[8\left(N+\frac34\right)\right]^{3/2}}
\quad{\rm and}\quad
n_s=1-\frac32\left(N+\frac34\right)^{-1}
\end{equation}

\begin{center}
\begin{figure}

\vspace{8cm}

\begin{picture}(10,10)
\put(-50,-500){
\leavevmode
\hbox{\epsfxsize=8in
\epsffile{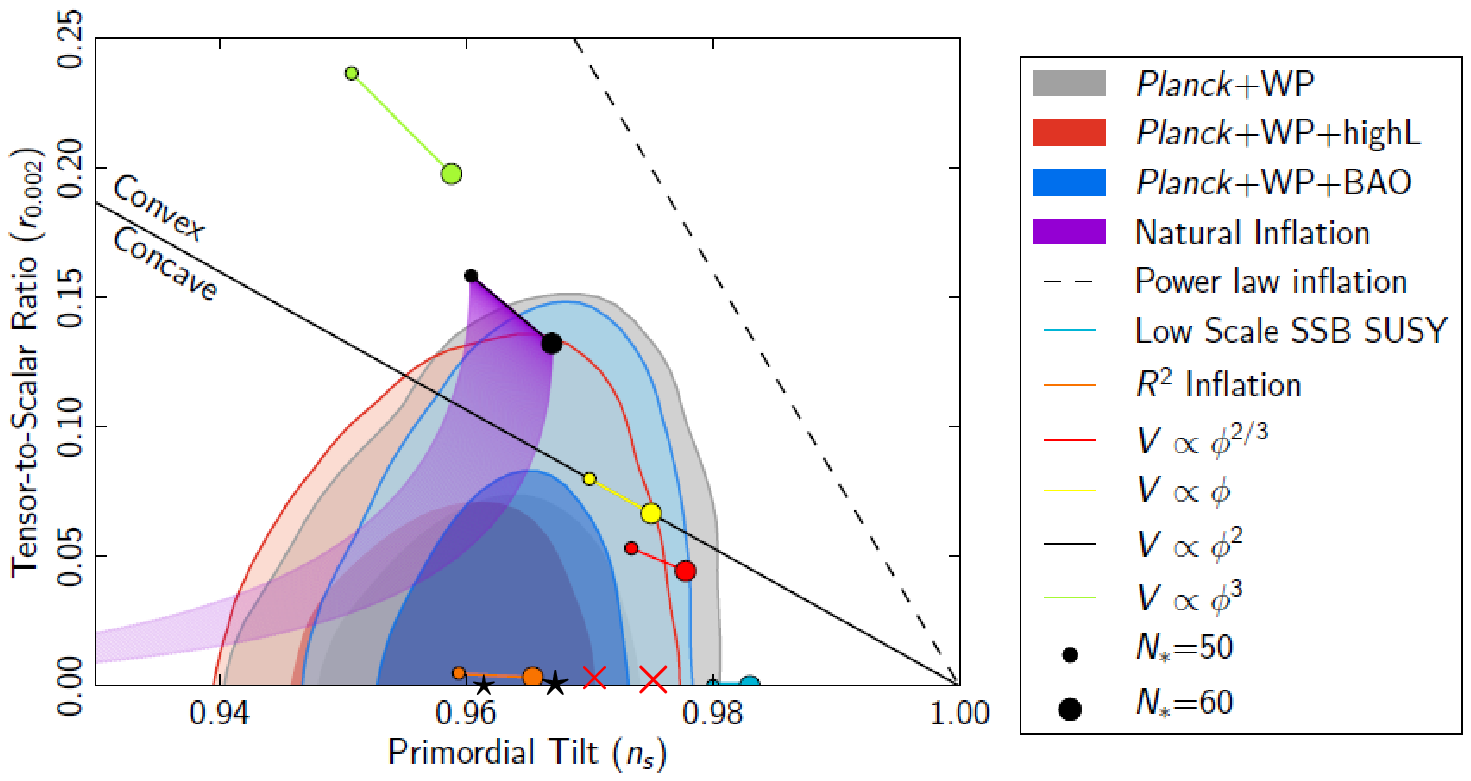}}
}
\end{picture}
\caption{\footnotesize
Shaft inflation for \mbox{$n=2$} is depicted with the large \{small\} cross
for \mbox{$N\simeq 60$} \{\mbox{$N\simeq 50$}\}. Shaft inflation for 
\mbox{$n\gg 1$} is depicted with the large \{small\} star for 
\mbox{$N\simeq 60$} \{\mbox{$N\simeq 50$}\}.
The difference with $R^2$-inflation (orange dots) stems from taking into account
the contribution of $\phi_{\rm end}$ so that \mbox{$N\rightarrow N+1$} in 
Eq.~(\ref{nbig}). Intermediate values of $n$ lie in-between the depicted points.
As evident, there is excellent agreement with the Planck observations.}
\label{fig2}

\end{figure}
\end{center}

For the moment, let us ignore the BICEP2 results and try to satisfy the 
Planck observations only.
Assuming \mbox{$\alpha\simeq 1$}, for \mbox{$N\simeq 60$} 
\{\mbox{$N\simeq 50$}\} we readily obtain \mbox{$n_s=0.975$} and 
\mbox{$r=2.99\times 10^{-3}$} \{\mbox{$n_s=0.970$} and 
\mbox{$r=3.91\times 10^{-3}$}\}. As shown in Fig.~\ref{fig2}, these values fall 
within the 95\% \{68\%\} c.l. contour of the Planck observations. 
Things improve further if we enlarge $n$.

Indeed, in the limit \mbox{$n\gg 1$}
Eqs.~(\ref{r}) and (\ref{nsN}) become
\begin{equation}
r=\frac{8\alpha^2}{n^2(N+1)^2}\rightarrow 0
\quad{\rm and}\quad
n_s=1-\frac{2}{N+1}\,.
\label{nbig}
\end{equation}
The spectral index is now the same as in the original $R^2$~inflation model
\cite{R2} (also in Higgs inflation \cite{Higgs}), which is not surprising since
we expect power-law behaviour to approach the exponential when 
\mbox{$n\rightarrow\infty$}. Now, 
for \mbox{$N\simeq 60$} \{\mbox{$N\simeq 50$}\} we obtain \mbox{$n_s=0.967$} 
\{\mbox{$n_s=0.961$}\}, which is very close to the best fit point for the 
Planck data, as shown in Fig.~\ref{fig2}.

\vspace{-1cm}

Now, let us incorporate in our thinking the BICEP2 results, which suggest that 
\mbox{$r=0.20\pm0.07$} \cite{bicep2}. From Eq.~(\ref{r}) it is readily seen 
that \mbox{$r\propto\alpha^{2n/(n+2)}$}. This means that the tensor production 
can be enhanced if the shaft is appropriately widened, i.e. if
$m$ is somewhat larger than $m_P$ without affecting the 
scalar spectral index, as seen in Eq.~(\ref{nsN}). Indeed, it is easy to show
that \mbox{$\alpha\simeq 50$} is enough to boost the tensor signal up to BICEP2 
values. For example, assuming \mbox{$N\simeq 50$} and \mbox{$\alpha=50$}, 
Eqs.~(\ref{r}) and (\ref{nsN}) give Table~1:

\begin{center}
\begin{tabular}{|c||c|c|}\hline
$n$ & $r$ & $n_s$ \\\hline\hline
2 & 0.200 & 0.970\\\hline
4 & 0.199 & 0.967\\\hline
6 & 0.141 & 0.966\\\hline
8 & 0.098 & 0.965\\\hline
\end{tabular}

\medskip

{\footnotesize
Table 1: Values of $r$ and $n_s$ for shaft inflation with \mbox{$N=\alpha=50$}
and \mbox{$n=2,4,6$} and 8.}
\end{center}

\begin{center}
\begin{figure}

\vspace{6cm}

\begin{picture}(10,10)
\put(50,-150){
\leavevmode
\hbox{\epsfxsize=5in
\epsffile{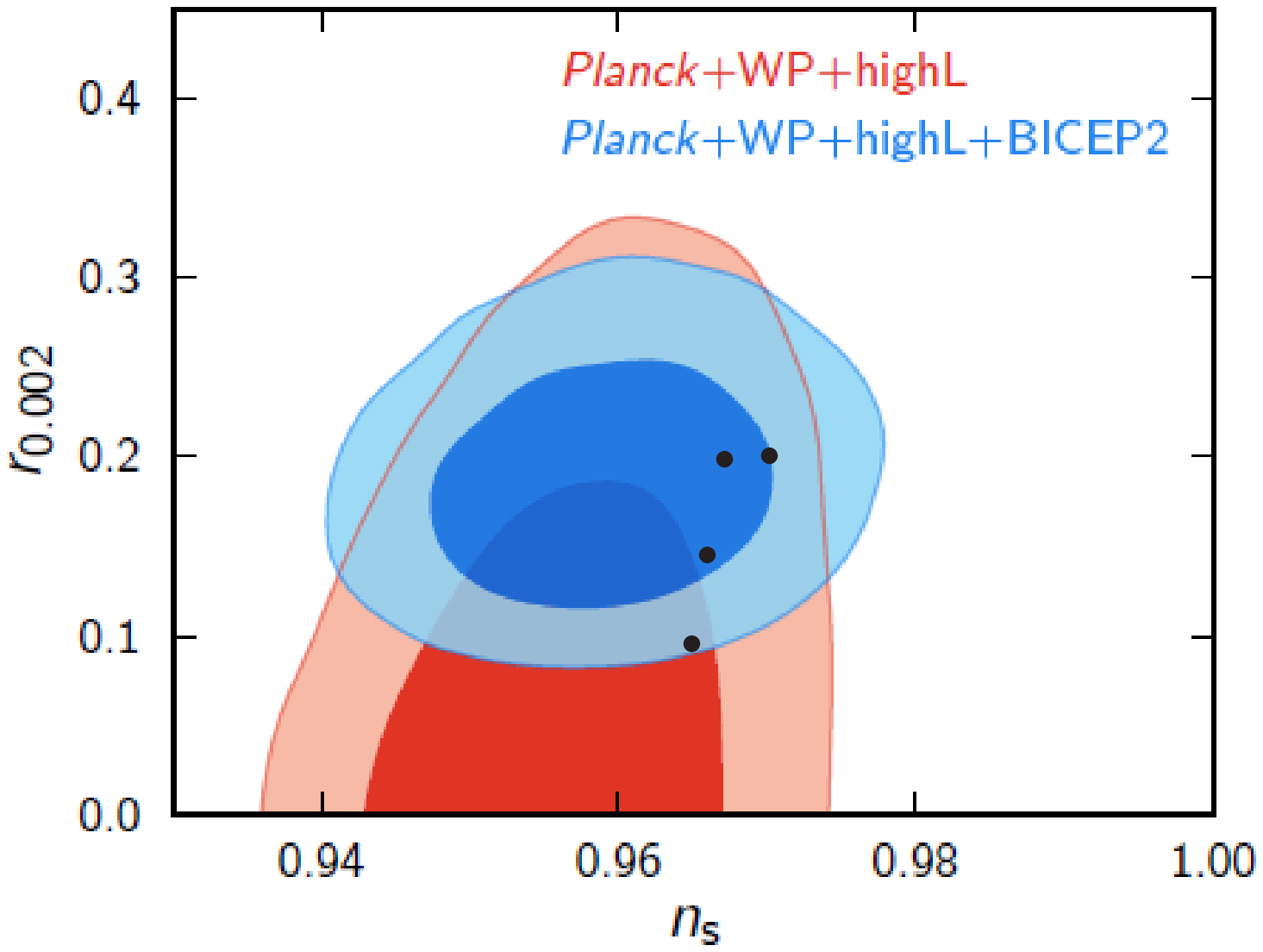}}
}
\end{picture}
\caption{\footnotesize
Allowing for a running spectral index, shaft inflation can accommodate the
BICEP2 results. The bullets depict the predictions of shaft inflation for 
\mbox{$n=2,4,6$} and 8 from right to left respectively. As evident, there is 
excellent agreement with the observations.}
\label{fig3}

\end{figure}
\end{center}

\vspace{-1cm}

Allowing for a running spectral index the BICEP2 results suggest the blue 
contours shown in Fig.~\ref{fig3}. However, it is easy to show that 
\begin{equation}
\frac{{\rm d}n_s}{{\rm d}\ln k}=
-\frac{2\left(\frac{n+1}{n+2}\right)}{\left(N+\frac{n+1}{n+2}\right)^2}\sim
-\frac{2}{N^2},
\end{equation}
which gives \mbox{$\frac{{\rm d}n_s}{{\rm d}\ln k}\sim 10^{-3}$} for 
\mbox{$N\simeq 50$}, so the running is not substantial.


An estimate for the required value of $M$ is obtained enforcing the COBE bound 
onto the curvature perturbation. Using Eqs.~(\ref{V}) and (\ref{fN}) we find
\begin{equation}
\sqrt{\calp_\zeta}\!=\!\frac{1}{2\sqrt 3\pi}\frac{V^{3/2}}{m_P^3|V'|}\Rightarrow
\!\left(\frac{M}{m_P}\right)^2\!\!=\!4\sqrt 3(n-1)\alpha^{-\frac{n}{n+2}}
\pi\sqrt{\calp_\zeta}\left[2(n\!-\!1)(n\!+\!2)\!\left(N+\frac{n+1}{n+2}\right)
\right]^{-\frac{n+1}{n+2}}.
\label{M}
\end{equation}
For \mbox{$n=2$}, \mbox{$N=60$} and \mbox{$\alpha=1$} \{\mbox{$\alpha=50$}\} 
and taking \mbox{$\sqrt{\calp_\zeta}=4.706\times 10^{-5}$} we get 
\mbox{$M=7.7\times 10^{15}\,$GeV} \{\mbox{$M=1.6\times 10^{15}\,$GeV}\},
which is close to the scale of grand unification, as expected.

Provided $\phi$ can be arbitrarily large [the vacuum density for large $\phi$
is constant and remains sub-Planckian, since \mbox{$M\ll m_P$} and 
\mbox{$V(\phi\gg m)\simeq M^4$}] one can show that slow-roll inflation can last 
for a huge number of e-folds. However, far away from the shaft, the potential 
becomes so flat that the inflaton finds itself in the so-called quantum 
diffusion zone, leading to eternal inflation \cite{eternal}. The criterion is 
as follows.

For eternal inflation to occur, the classical variation of the inflaton field
$|\dot\phi|$ needs to become subdominant to the quantum variation of $\phi$, 
which is given by the Hawking temperature \mbox{$\delta\phi=H/2\pi$} per
Hubble time \mbox{$\delta t=H^{-1}$}. Comparing the two it is easy to show
that
\begin{equation}
|\dot\phi|
\mbox{\raisebox{-.9ex}{~$\stackrel{\mbox{$>$}}{<}$~}}
\frac{\delta\phi}{\delta t}\Leftrightarrow|V'|
\mbox{\raisebox{-.9ex}{~$\stackrel{\mbox{$>$}}{<}$~}}
\frac{3}{2\pi}H^3,
\end{equation}
where we used the slow-roll equation of motion \mbox{$3H\dot\phi\simeq-V'$}.
In view of Eq.~(\ref{V}) and using the Friedman equation 
\mbox{$V(\phi)\simeq 3(Hm_P)^2$}, after some algebra, one can show
\begin{equation}
|\dot\phi|
\mbox{\raisebox{-.9ex}{~$\stackrel{\mbox{$>$}}{<}$~}}
\frac{\delta\phi}{\delta t}\Leftrightarrow
\frac{N}{N_*}
\simeq\frac{N+\frac{n+1}{n+2}}{N_*+\frac{n+1}{n+2}}
\mbox{\raisebox{-.9ex}{~$\stackrel{\mbox{$<$}}{>}$~}}
\left(\sqrt{\calp_\zeta}\right)^{-\frac{n+2}{n+1}}\sim 10^{4-6},
\end{equation}
where we also used Eq.~(\ref{M}), we considered that
\mbox{$1<\frac{n+2}{n+1}<\frac32$} and with $N_*$ we have denoted the remaining
e-folds, when the cosmological scales leave the horizon, i.e. 
\mbox{$N_*\simeq 60$}. 

Thus, we see that, even though the multiverse may be undergoing eternal 
inflation, our region finds itself relatively close to the potential shaft such
that slow-roll takes over and the inflaton gradually moves into the shaft. The 
inflaton slow-rolls for a few 
millions of e-folds before the cosmological scales exit the horizon and 
\mbox{$N_*\simeq 60$} after this. Eventually, inflation, in our region, ends 
and the inflaton oscillates at the bottom of the shaft, leading to (p)reheating
and the hot big bang. Meanwhile, elsewhere in the multiverse, eternal inflation 
continues. One can imagine that there may be a large number of shafts 
puncturing the inflationary plateau (leading to different vacua, possibly with 
different values of $n$). 
Some regions of the multiverse are close to the shafts in such a way that 
eternal inflation is superseded by classical slow-roll which attracts the 
system into the shaft in question. Our observable universe is such a case.

In summary, we introduced and studied a new family of inflation models, which 
we called shaft inflation. The models correspond to the scalar potential given 
in Eq.~(\ref{V}) and are parametrised by \mbox{$n>1$}. We obtained the models 
in the context of global supersymmetry starting with a superpotential, which
interpolates from a generic monomial to an O'Raifearteagh form for small to 
large values of the inflaton field respectively. However, shaft inflation can 
be obtained in different setups, as mentioned, for example,  in the case 
\mbox{$n=2$}. We showed that we obtain values for the spectral index $n_s$ or 
the tensor to scalar ratio $r$ that are in excellent agreement with the latest 
observations of the Planck satellite and the BICEP2.

\bigskip

\noindent
{\bf Acknowledgements}

\medskip

\noindent
KD is supported (in part) by the Lancaster-Manchester-Sheffield Consortium for 
Fundamental Physics under STFC grant ST/J000418/1.

%
%
%

\end{document}